# *Transient Phenomena of Mercury*

**Oleksiy Arkhypov, PhD**

Oleksiy.Arkhypov@oeaw.ac.at

**Abstract**

The similarity of Mercury to the Moon raises the question on mercurian analogues of lunar transient phenomena. However, since the mid-1960s, this topic has remained out of attention. Nevertheless, the discoveries of the 21st century and the *BepiColombo* mission give hope for the registration of Mercury transient phenomena. In this regard, it is shown that there are forgotten earth-based observations of anomalous phenomena on Mercury, which now fit into modern knowledge about this planet. Therefore, ALPO could contribute to this mysterious aspect of space exploration.

## 1. Introduction

The topic of variable clouds and other atmospheric effects (limb haze, polarization anomalies at horns) on Mercury was intensively discussed in the astronomical press of the 19th and first half of the 20th centuries (e.g., Sandner, 1963). However, after discovery of the real planet's rotation period in 1965, it became clear that many previous reports of such Mercurian changes were the result of erroneous beliefs about Mercury's rotation (Colombo and Shapiro, 1966). Mercury doesn't have enough mass to hold an atmosphere (do not confuse with exosphere). Therefore, observations of changes that contradict theoretical expectations were ignored and forgotten.

However, the low mass does not prevent the Moon from having a temporary local atmosphere and dust clouds as a result of outgassing of its interior or meteoroid impacts. Such manifestations of non-stationary processes are known as *lunar transient phenomena* (Crotts, 2008). In particular, such outgassing is confirmed in situ during the Japanese *SELENE/Kaguya* mission (Ohtake et al. 2024). The lunar dust fountains were discovered in the data of the *Lunar Atmosphere and Dust Environment Explorer* (LADEE) (Xie et al. 2020).

Research has revealed a striking similarity between surfaces of Mercury and the Moon (Strom, 1974). Accordingly, one should expect *Mercury Transient Phenomena* (MTPs) similar to the lunar case. The purpose of this work is to search among forgotten publications for reports on probable MTPs. Such reports have been forgotten as rare exotics that contradicted the prevailing ideas about Mercury.

However, the accumulation of new knowledge has now made possible what was once considered unbelievable. The finds could inspire members of the ALPO to study "white spots" of Mercurian science.

## 2. New Knowledge

Earth-based observations have revealed high variability in Mercury's exosphere, reminiscent of the lunar case (Leblanc et al. 2022). This variability is an indicator of transient phenomena on the surface of the planet. Thus, the *Fast-Imaging Plasma Spectrometer* on board of the MESSENGER spacecraft detected the group of sodium ions that were recently ionized. This "***transient enhancement** of Mercury's exosphere*" was observed at extremely high altitudes of 5300 km (Jasinski et al. 2020). It has been argued that the global exosphere cannot explain these high-altitude observations indicating micrometeoroid impact. In general, seasonal variations and transient enhancements in the exosphere of Mercury were attributed to such impacts.

Of course, in addition to micrometeoroids, big impacts also occur, as evidenced by meteorites falling to Earth from Mercury (Beatty, 2012). Another result of such impacts is a ring of dust along Mercury's orbit (Pokorný et al. 2023).

As a result, the planet is surrounded by a variable dust cloud of impact ejecta (Krüger et al. 2024). This near-Mercury dust is electrically charged and can levitate, just like above the Moon. However, the fluxes of electrifying factors (solar ultraviolet radiation and wind plasma) on Mercury are approximately an order of magnitude higher. This could generate larger dust clouds than those observed over the Moon.

In addition to impacts, dust can be lifted by jets of gas gushing from under the surface of Mercury. Such outgassing has been detected on the Moon as volatile emanations from interiors (Lawson et al. 2005) and from polar ice (Ohtake et al. 2024). Mercury, with its discovered paleo-volcanism and polar ice, is considered an arena of intense outgassing (Deutsch and Head, 2018).

In summary, the current knowledge portrays Mercury as a planet favorable to various transient phenomena.

## 3. Forgotten pre-cursors

Could watchful observers notice some of the phenomena described above? Libraries keep many interesting but forgotten hints about this. Let's consider them in the order of the expected sequence of events.

### 3.1. Interior or impact outgassing

Intriguing words are found in the monography of a soviet planetary expert, professor Vsekhsvyatskii: "*Herschel observed flashes on the surface of Mercury and spoke directly about active volcanoes on this planet*" (Vsekhsvyatskii, 1967). Unfortunately, no reference was made and I cannot find the source of this statement.

A forgotten Soviet brochure from 1963 reported: "*Due to volcanism, gases and perhaps water vapor are released on Mercury*" (Nesterovich, 1963, p. 31). Really, measurements of the polarization of light reflected from Mercury in the twilight region in April, September and October 1950 led the famous O. Dollfus to the conclusion: "*the distribution of polarization along the terminator is that predicted by the existence of a weak atmosphere*" (Dollfus, 1957). Moreover, he noted the abnormal, transient polarization events. In particular, "*on October 5, 1950, under the phase angle V = 78°.5, the **North Pole** stood out with a polarization that was 6 thousandths too low … The next day, October 8, … the polarimeter indicated a value that was too low at the **North Pole**, around 7 thousandths*" (Dollfus, 1957, p. 44). In 1965 the main Soviet astronomical journal reported the spectrographic recording of a transient atmosphere of volcanic $CO_2$ gas on Mercury on October 14, 1963 (Moroz, 1965). On April 24 and May 3, 1963, the transient presence of hydrogen atmosphere on Mercury was recorded spectrographically (Kozyrev, 1964).

These statements looked like heresy in the 1960s-1980s. However, in 1991 the frozen water was discovered at **North Pole** of Mercury (Beatty, 1991). Following confirmation by ground-based radar and the MESSENGER spacecraft (Glanzberg et al. 2023), polar $H_2O$ ice now appears to be a legitimate source of a transient hydrogen atmosphere. Moreover, the *MESSENGER*'s discovery of sublimation hollows, and chaotic terrains on the Mercury surface is considered as an evidence of global volatile-rich layers extending several kilometers in depth (Rodriguez et al. 2023). In this regard, experts admit relatively recent manifestations of volcanism on Mercury.

Therefore, now the old reports about the outgassing and transient atmosphere of Mercury do not contradict the modern science vision.

### 3.2. Dust plumes

Meteoroid's impacts and outgassing are known drivers of dust lifting in the Moon (Crotts, 2008; Ohtake et al. 2024) as well as in Mercury (Krüger et al. 2024). Therefore, dust plumes could be observed above the planet.

On March 31, 1800, J. H. Schröter recognized two small bright prominences, which projected outside the limb, very close to the end of the southern horn. He wrote: "*I noticed a pair of unequal, knotty or jagged places on the outer edge [of Mercury] in a and b [points in figure 12], which on a smaller scale resembled the Dörfel mountain*

4*range of the Moon, seemed to be similar*" (Schröter 1800, p. 63). However, the only available scan of the original source does not contain the indicated figure. There all tables have been scanned in an unexpanded state (!), demonstrating the inferiority of digital copies of rare publications. Fortunately, a copy of this image was found in another rarity from the home library (Figure 1).

Obviously, it was impossible to see real mountains above Mercury, since the diameter of the planet was only 8 arc seconds. Judging by Figure 1, the height of the observed above-limb irregularities was tens of kilometers. This estimate excludes mountains but suggests dust plumes. Comparison of Schröter's drawing with the view of Mercury on the same date (Figure 1) shows the absence of bright spots of the planetary surface in the region of the prominences. Consequently, Schröter's prominences could not be an irradiation illusion.

Sandner (1963, p. 34) mentioned that Vaughn saw a similar phenomenon on November 8, 1941.

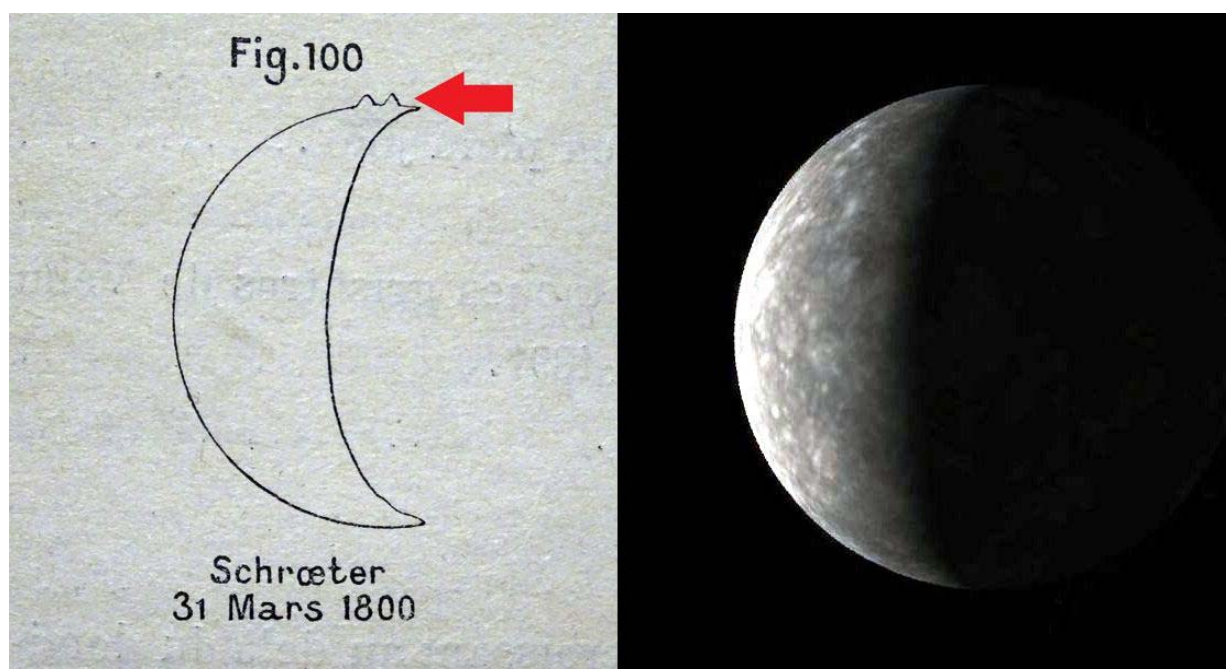

*Figure 1.* Protuberances seen by J. H. Schröter on March 31, 1800 (left; Trouvelot, 1892) are compared with the corresponding image of Mercury (right) generated by a virtual planetarium "Stellarium 1.2" for that date.

Figure 2a shows "*a small white, horn-like point*" above the limb of Mercury on July 17, 1885, which transformed into a cloud (Corliss 1985, p. 38-39). The observer notes: "*A few minutes later fancied to see a whitish patch deeper on bright limb*". This patch (cloud) was seen several days. The virtual planetarium does not show any bright spots of the planetary surface in the region of the phenomenon.

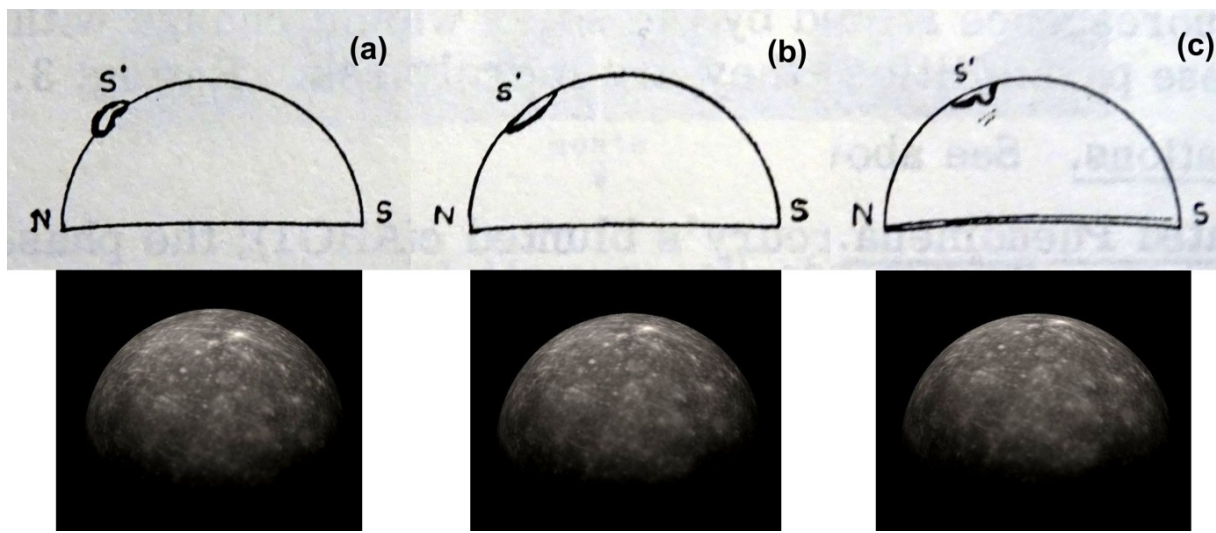

*Figure 2.* Drawings of the evolution of a white projection (labeled as s') above the Mercury's surface (Ballot, 1885): *(a)* the limb protrusion on July 17, 1885; *(b)* light patch on 07/19/1885; *(c)* the same cloud on 07/21/1885. The bottom row shows the view of Mercury generated by the virtual planetarium "Stellarium 1.2" for the corresponding dates.

### 3.3. Dust clouds

Raised dust forms clouds (Berezhnoy et al. 2019) and even a global envelope around the Moon (Horányi et al. 2015). Similarly, dust clouds could be observed on Mercury.

In particular, E.M. Antoniadi (1933) claimed to have repeatedly noticed whitish clouds near the limb of Mercury. His drawings (Figure 3 for example) depicts a *"brilliant whiteness" interpreted as* clouds (nuées) or haze (voile) where there are no high-albedo features on Mercury, but where the maximum optical thickness of the dusted layer would be expected.

Figure 4 shows the modern drawings of the sudden formation and disappearance of a light cloud around Mercury's north pole, where ice deposits were detected. Similar sudden (shorter 35 minutes) appearance of a northern-polar bright spot was recorded as CCD image (Figure 5).

The presented examples of transient phenomena on Mercury are not associated with the slow rotation of ordinary high-albedo regions, as comparison with synthetic images of the planet shows. Therefore, these anomalies deserve attention.

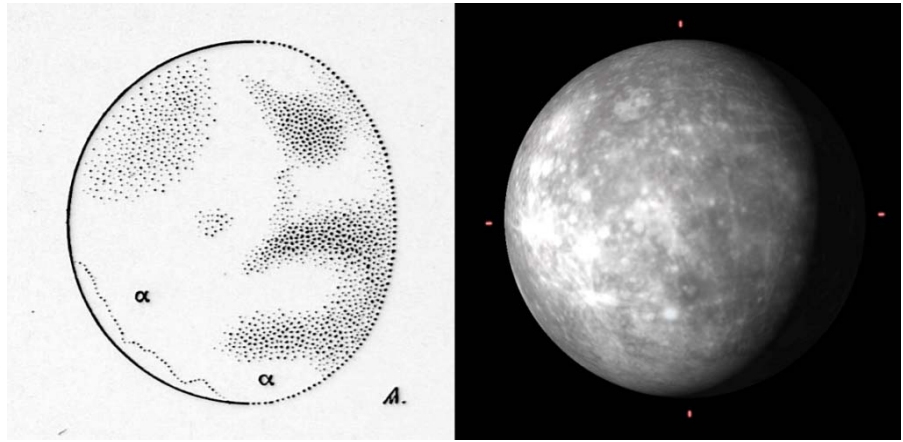

***Figure 3.*** *Whitish haze near the limb of Mercury (the dotted line in the left drawing; fig. 251 in Antoniadi, 1933) is compared with the planetary view, which is simulated using the virtual planetarium Stellarium 1.2 for the observation in Meudon on 1927/09/27, 17:35 UT (right).*

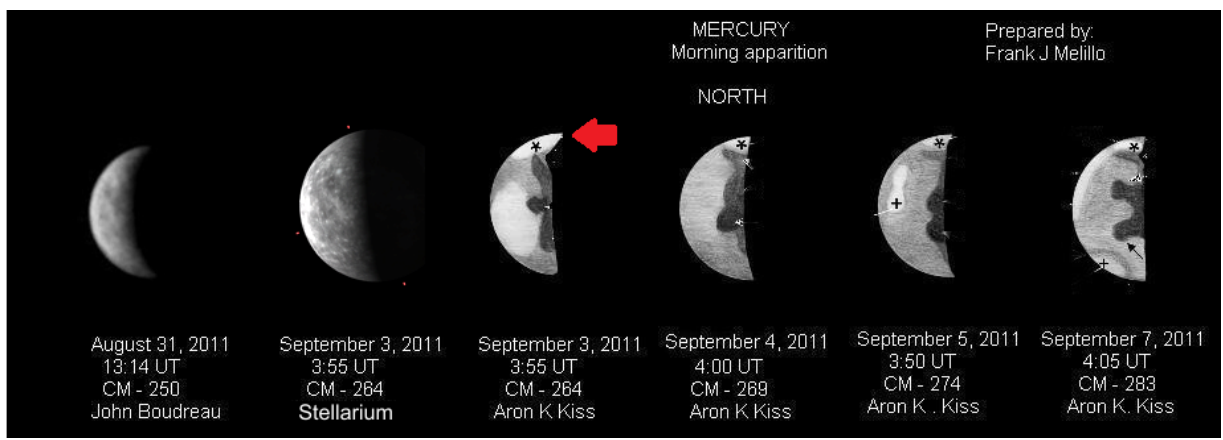

***Figure 4.*** *ALPO Archive demonstrates the sudden formation of bright cloud in the northern polar region on September 3, 2011 (arrowed). The second image from left is the author's simulation of the planetary view to compare with arrowed drawing (the virtual planetarium Stellarium 1.2 was used).*

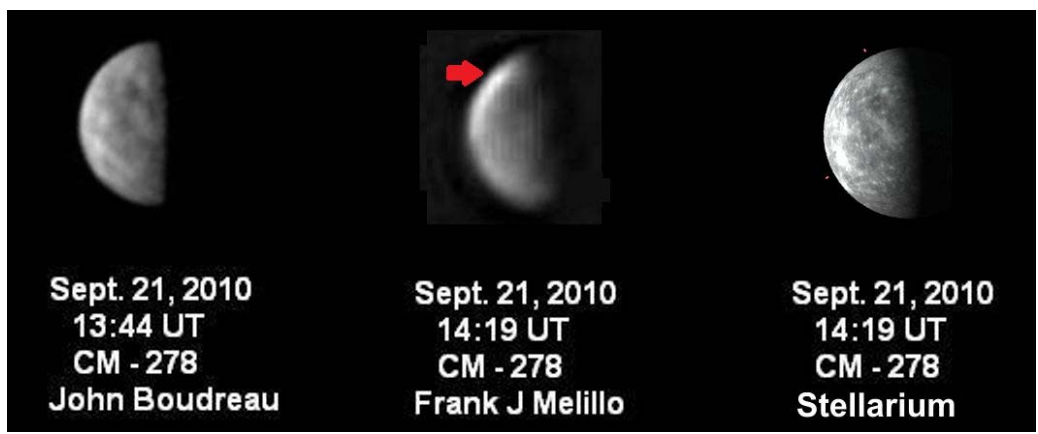

***Figure 5.*** *Bright cloud suddenly appeared in the northern polar region on September 21, 2010 (ALPO Archive). The author's simulation of the Mercury view is shown in right to compare.*

## 3.4. Global dust envelope

It is logical to assume that local dust clouds could feed the global dust envelope around Mercury. Variable envelope thickness can manifest itself in different ways.

With a small thickness, the dust layer can be seen as an analogue of the lunar horizon-glow (Rennilson and Criswell 1974). In this case, the dust could be seen as prolongations of planetary crescent horns due to the forward scattering of sunlight. This is the effect that was observed by R.W. Middleton on May 19, 1975 (Fig. 6a; or fig. 3 in Robinson, 1976).

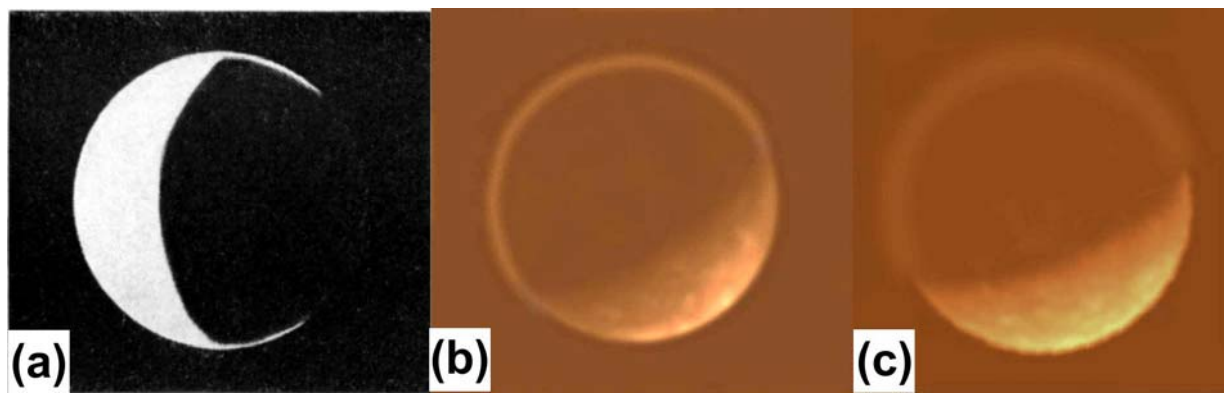

*Figure 6. Forward scattering of sunlight by the global dust envelope of Mercury: **(a)** the drawing from the BAA archive of the horn prolongation on May 19, 1975 (Robinson, 1976); **(b)** the annular Mercury on May 18, 1896 (Brenner, 1896) and 14 April, 1944 (Sandner, 1963, p. 52) in **(c)**. Panels (b) and (c) are the author's simulations using "Stellarium 1.2".*

If the thickness of the layer of lifted dust is sufficient to reach the sunlight along the entire night limb of the planet, then a light ring is formed. One can call this, so far unnamed, special observational phenomenon as an *Annular Mercury*. In particular, L. Brenner saw "*the dark side surrounded by an aureole*" on May 18, 1896 (Brenner, 1896). The light ring was seen as "very conspicuous" object by two observers despite of change of oculars as well as the position of the planet in field of view. A simulation of this phenomenon is shown in Figure 6b. Another experienced observer, K. Novák, saw the Annular Mercury on 14$^{th}$ April 1944 (Figre 6c; *Sandner, 1963, p. 52*).

Dollfus (1957) observed the transient global dust cloud at the Mercury limb: "*On October 7, 1950, for [the phase angle] V = 69°.5, the entire limb shows excessive polarization, especially through the green filter. The next day, October 8, I no longer observed anything abnormal at the limb. … These observations are difficult to make and therefore cannot provide certainty; however, they could reveal the existence of the transient hazes (de voiles passagers) suspended in the atmosphere, because a **cloud of dust** raised by the wind weakens the polarization at these viewing angles. Such observations deserve to be taken up and developed.*"

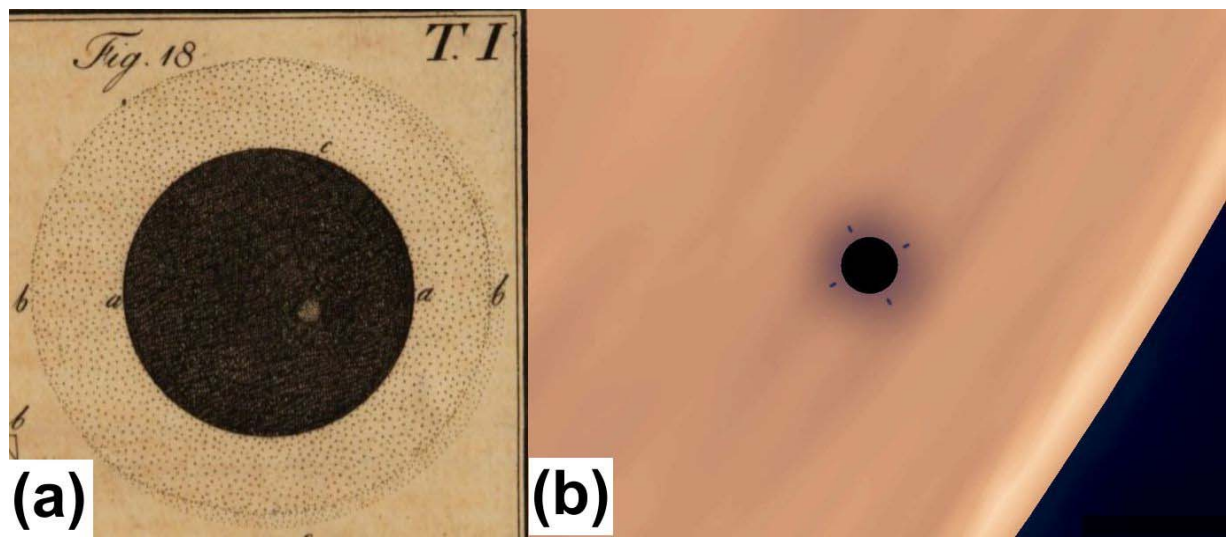

*Figure 7.* Dark halo of dust extinction around Mercury during its transits across the solar disk: *(a)* the original figure from the report describing the transit on May 7, 1799 (Schröter, 1800); *(b)* the author's simulation of a dark halo seen on May 5, 1832 (Moll, 1833).

Rarely, the shadow of a dust envelope in the form of a dark halo around the planet has been observed during Mercury's transits across the solar disk. Figure 7a shows the original sketch by J. H. Schröter (1800). This experienced observer described the transit on May 7, 1799:

"*As sharply defined as Mercury itself appeared, a ring of light was faintly visible all around it over a fairly considerable width, which was of a barely noticeably brighter color, at least of a different color, than the rest of the sun's surface, but of a somewhat **duller light**, and that ended all around in a blurred, barely noticeable grayish border; on the left or western side but, as far as I remember, of a slightly different width than on the eastern one. … I estimated its width to be 1/4 of Mercury's diameter or more.*" (Schröter, 1800, p. 24, 30).

His assistant, Mr. Harding, confirmed this observation using another telescope. The dark ring around Mercury was observed throughout the whole transit in various telescopes and eyepieces, therefore it was real, although difficult to see. Schröter saw the dark ring sing 13-foot telescope having a mirror of D=24 cm (9.5 inch) in diameter. Consequently, its resolution limit according to the Rayleigh criterion was about 1.22 $\lambda/D$ = 0.6 arcseconds, and the thickness of the dark ring was 12.09/4 = 3.02 arcseconds, or 5 times the found diffraction limit of the telescope. Therefore, the observed dark halo around Mercury cannot be a diffraction fringe that would be much narrower and bright, no dark.

The bright diffraction fringe or the "bright ring" in Corliss (1985, p. 44) around transiting Mercury as well as at the lunar limb were noted numerously (Clerke, 1902). However, after Schröter and Harding, the dark halo was seen only on May 5, 1832:

*"It occurred to us that the Sun's disc, seen with different colored glasses, seemed steadily darker around Mercury, moving more towards the violet. The darkness continued to descend from the planet undiminished, and lost itself in the bright shine of the Sun. Mr. Rueb believed that the dark nebula was not entirely concentric with Mercury, but extended further on the upper left side of the planet, in the inverted field of view."* (Moll, 1832)

Consequently, in each case described, the dark halo was observed simultaneously by two observers using different telescopes. Thus, the dark halo turned out to be a real, but extremely rare phenomenon.

## 4. Conclusion

Abovementioned old reports clearly show that Mercury is not a dead planet, but rather a changeable world. However, changes on it, that are noticeable from the Earth, occur very rarely on a scale of decades. Correspondingly, a few time-limited space missions cannot give representative information on such MTPs. However, an earth-based Mercury patrol has certain advantages on long timescale. It's especially important to keep an eye out for sudden dust ejections that could threaten orbiters like the *BepiColombo* probes. The second task could be the detection and identification of active regions on Mercury. In this regard, the capabilities of ALPO cannot be overestimated.

**References**

Antoniadi E.-M. (1933). "La planete Mercure". *L'Astronomie*, Vol. 47, No. 12, pp. 545-558.

Ballot J. (1885). "Suspected white markings on Mercury". *English Mechanic*, Vol. 42, pp. 139, 199.

Beatty J.K. (1991). "Mercury's cool surprise". Sky and Telescope, Vol. 83, No. 1, pp. 35-36.

Beatty J.K (2012). "Mercury's Marvels". *Sky and Telescope*, Vol. 123, No. 4, pp. 26-33.

Berezhnoy A. A., Velikodsky Yu. I., Zubko E. et al. (2019). "Detection of impact-produced dust clouds near the lunar terminator". *Planetary and Space Science*, Vol. 177, id. 104689.

Brenner L. (1896). "Visibility of the dark side of Mercury". *Journal of the British Astronomical Association*, Vol. 6, No. 8, p. 387.

Colombo G. and Shapiro I.I. (1966). "The rotation of the planet Mercury". *The Astrophysical Journal*, Vol. 145, No. 1, pp. 296-307.

Corliss W.R. (1985). *The Moon and the planets: a catalog of astronomical anomalies*. Glen Arm: The Sourcebook Project.

Crotts A.P.S. (2008). "Lunar Outgassing, Transient Phenomena, and the Return to the Moon. I. Existing Data". *The Astrophysical Journal*, Vol. 687, No. 1, pp. 692-705.

Deutsch A. N. and Head J. W. (2018). "Production Function of Outgassed Volatiles on Mercury: Implications for Polar Volatiles on Mercury and the Moon". *Mercury: Current and Future Science of the Innermost Planet, Proceedings of the conference held 1-3 May, 2018 in Columbia, Maryland. LPI Contribution*, No. 2047, id.6121.

Dollfus A. (1957). "Étude des planètes par la polarisation de leur lumière". *Supplements aux Annales d'Astrophysique*, Vol. 4, pp. 42, 44.

Glantzberg A.K., Chabot N.L., Barker M.K. et al. (2023). "Investigating the Stability and Distribution of Surface Ice in Mercury's Northernmost Craters". *The Planetary Science Journal*, Vol. 4, No. 6, id.107.

Horányi M., Szalay J. R., Kempf S. et al. (2015). "A permanent, asymmetric dust cloud around the Moon". *Nature*, Vol. 522, No. 7556, pp. 324-326.

Jasinski J. M., Regoli L., Cassidy T. et al. (2020). "A transient enhancement of Mercury's exosphere at extremely high altitudes inferred from pickup ions". *Nature Communications,* Vol. 11*,* id. 4350.

Kozyrev N.A. (1964). "The atmospere of Mercury". *Sky and Telescope*, Vol. , No. 6, pp. 339-341.

Krüger H., Thompson M.S., Kobayashi M. et al. (2024). "Understanding the dust environment at Mercury: from surface to exosphere". *The Planetary Science Journal*, Vol. 5, No. 2, id. 36.

Lawson S.L., Feldman W.C., Lawrence D.J. et al. (2005). "Recent outgassing from the lunar surface: The Lunar Prospector Alpha Particle Spectrometer". *Journal of Geophysical Research*, Vol. 110, No. E9, id. E09009.

Leblanc F., Schmidt C., Mangano V. et al. (2022). "Comparative Na and K Mercury and Moon Exospheres". *Space Science Reviews*, Vol. 218, No. 1, id. 2.

Moll G. (1832). "Schreiben des Herrn Professors und Ritters Moll an den Herausgeber". *Astronomische Nachrichten*, vol. 10, No. 14(229), pp. 201-206.

Moroz V. I. (1965). "Infrared Spectrum of Mercury (λ=1.0-3.9μ)". *Soviet Astronomy*, Vol. 8, No. 6, pp. 882-889.

Nesterovich E.I. (1963). *Merkuriy* (Mercury). Moscow: Znaniye, p. 27-33.

Ohtake M., Nakauchi Y., Tanaka S. et al. (2024). "Plumes of water ice/gas mixtures observed in the lunar polar region". *The Astrophysical Journal*, Vol. 963, No. 2, id.124.

Pokorný P., Deutsch A.N., and Kuchner M.J. (2023). "Mercury's Circumsolar Dust Ring as an Imprint of a Recent Impact". *The Planetary Science Journal*, Vol. 4, No. 2, id.33.

Rennilson J.J. and Criswell D.R. (1974). "Surveyor Observations of Lunar Horizon-Glow". *The Moon*, Vol. 10, No. 2, pp. 121-142.

Robinson J.H. (1976). "Report on the planet Mercury". *Journal of the British Astronomical Association*, Vol. 86, No. 10, pp. 487 - 490.

Rodriguez J.A.P., Domingue D., Travis B. et al. (2023). "Mercury's Hidden Past: Revealing a Volatile-dominated Layer through Glacier-like Features and Chaotic Terrains". *The Planetary Science Journal*, Vol. 4, No. 11, id.219.

Sandner W. *The Planet Mercury*. London: Faber and Faber, 1963.

Schröter J.H. (1800). *Beiträge zu den neuesten astronomischen Entdeckungen*, Vol. 3, Göttingen: Commission der Vandenhoek-Ruprechtischen Buchhandlung.

Strom R.G. (1974). "The planet Mercury as viewed by Mariner 10". *Sky and Telescope*, Vol. 47, No. 6, pp. 360-369.

Trouvelot M.E.L. (1892). "Observations sur les planètes Vénus et Mercure". *Bulletin de la Société Astronomique de France*, Vol. 6, No. 3, pp. 61-147.

Xie L., Zhang X., Li L. et al. (2020). "Lunar dust fountain observed near twilight craters". *Geophysical Research Letters*, Vol. 47, No. 23, id. e89593.

Vsekhsvyatskii S.K. (1967). *Priroda i proiskhozhdeniye komet i meteornogo veshchestva* (The nature and origin of comets and meteoric matter). Moscow: Prosveshcheniye, p. 130.